# Translational Spacetime Symmetries in Gravitational Theories

**R. J. Petti**

The MathWorks, Inc., 3 Apple Hill Drive, Natick, MA 01760, U.S.A.

E-mail: rjpetti@alum.mit.edu

**Abstract**
How to include spacetime translations in fibre bundle gauge theories has been a subject of controversy, because spacetime symmetries are not internal symmetries of the bundle structure group. The standard method for including affine symmetry in differential geometry is to define a Cartan connection on an affine bundle over spacetime. This is equivalent to (1) defining an affine connection on the affine bundle, (2) defining a zero section on the associated affine vector bundle, and (3) using the affine connection and the zero section to define an 'associated solder form,' whose lift to a tensorial form on the frame bundle becomes the solder form. The zero section reduces the affine bundle to a linear bundle and splits the affine connection into translational and homogeneous parts; however it violates translational equivariance / gauge symmetry. This is the natural geometric framework for Einstein-Cartan theory as an affine theory of gravitation. The last section discusses some alternative approaches that claim to preserve translational gauge symmetry.

PACS numbers: 02.40.Hw, 04.20.Fy

## 1. Introduction

Since the beginning of the twentieth century, much of the foundations of physics has been interpreted in terms of the geometry of connections and curvature. Connections appear in three main areas.

1) *Affine geometry of spacetime*. General relativity (GR) is based on a curved Riemannian manifold of mixed signature. Einstein-Cartan theory (EC) introduces affine torsion to extend GR to include spin.

2) *Gauge theories of internal symmetries*. Electromagnetic theory (Weyl 1929), strong interactions (Yang and Mills 1954, Ne'eman 1961, Gell-Mann 1962, Gell-Mann and Ne'eman 1964) are described by connections on unitary bundles. Later strong, weak and electromagnetic forces were combined in a larger unitary gauge theory.

3) *Hamiltonian and Lagrangian mechanics*. Hamiltonian and Lagrangian mechanics are based on "symplectic geometry," which is based on a connection on a complex line bundle over phase space. (Abraham and Marsden 1994, Guillemin and Sternberg 1984).

Table 1 summarizes many of the applications of curvature in basic physics.

**Table 1:** Some applications of connections and curvature in physical theories

| Continuous Differential Geometry | Fundamental Field Theories | Discrete Lattices (order defects) | Continuous Models of Materials |
|---|---|---|---|
| Linear (rotational) curvature (Riemannian curvature in metric case) | Gravitational force in general relativity & Einstein-Cartan theory | Disclinations (line defects of rotational order) | Incompatible strain |
| Translational curvature (affine torsion) | Spin-spin contact force (very small) in Einstein-Cartan theory | Dislocations (line defects of transla-tional order) | Dislocation density |
| Generalized curvature (curvature on fibre bundles) | Yang-Mills forces (electromagnetic, weak, strong forces) | Dispirations (line defects of internal order) | Histeresis (e.g. elasto-plastic, magnetic) |
| Symplectic structure (on phase space) | Hamiltonian and Lagrangian mechanics | | |
| Extrinsic curvature (of embedded surfaces) | • Initial value problem in general relativity<br>• Equations of classical string theory | | • Surface tension<br>• Curvature parameter 'R' in hard-body models of thermodynamics of fluids |

Some common themes emerge in physical theories using connections on fibre bundles. (Petti 1976, 2001, Kleinert 1987). Fibre spaces represent idealized highly symmetric local models. Connections show how to connect local domains of high symmetry along a path. Curvature tensors describe defects in the way the highly symmetric local domains fit together.[1] Bianchi identities express defect conservation constraints.

In a more perfect part of the universe, this paper would be unnecessary. Unfortunately, in our neighborhood, the controversy over how to include spacetime translations in gauge theories has persisted for 50 years (Utiyama 1956, Kibble 1961, Sciama 1962, Trautman 1973, Petti 1976, 2001, Ne'eman 1979, Lord and Goswami 1985, 1986, Lord 1976, Lord and Goswami 1988, Hehl et al. 1995, Blagojevic 2002, Tresguerres 2002). Authors disagree on whether to introduce translational symmetry through Cartan connections—which violate translational equivariance / gauge symmetry—or to reject Cartan connections in an effort to preserve full translational gauge symmetry.

This paper aims to resolve this controversy by showing that the Cartan connection approach is correct, and that the proposed alternatives also violate translational equivariance. We also introduce a new approach to the relationship between affine and (homogeneous) linear connections based on choice of a zero section of the associated affine vector bundle. The key issues are (1) equivariance / gauge invariance of translational symmetries requires that all computations can be expressed using any affine basis in each fibre of the associated bundle; and (2) using in a variational principle the 1-1 tensor field that identifies base space and fibre tangents requires fixing the translational gauge, that is, requires choosing a fixed origin point in each affine fibre space.

---

[1]    In continuum models of crystals, fibre spaces represent the idealized perfect crystals, affine torsion represents dislocations, linear (rotational) curvature represents disclinations, and gauge curvature represents dispirations.

Section 2 prescribes a procedure for including translational symmetry in theories based on connections (Petti 1976, 1986, 2001).[2] Section 3 explores the equivalent formulation in terms of Cartan connections of differential geometry. Section 4 rebuts criticisms of this approach and diagnoses errors that have appeared in the literature. Appendix A reviews the basic definitions of equivariance / gauge invariance. Appendix B describes how affine connections and the zero section combine to define the 'associated solder form' and thereby the solder form.

## 2. Procedure for Including Spacetime Translations

Notation:

- $\Xi$ = a (base) manifold of dimension $n$. Let $\xi^\mu$ be a local coordinate system on $\Xi$.

- $P(\Xi, G)$ = a principal fibre bundle over $\Xi$ with structure Lie group $G$ and bundle projection $\pi : P \rightarrow \Xi$. Denote the Lie algebra of $G$ by $L(G)$.

- $\omega : TP(\Xi, G) \rightarrow L(G)$ is a connection form.

- $X$ = differentiable manifold on which $G$ acts effectively[3] as a left transformation group. $X$ need not be a linear or affine space. $L(G)$ is represented as vector fields on $X$, denoted $L(G) \cdot X$. Let $x^i$ be smooth coordinates on $X$.

- $B(\Xi, G, P, X)$ = the associated fibre bundle (associated with $P$) with fibre space $X$.

### 2.1. Define Affine Fibre Bundles

Let $P(\Xi, A(n))$ be a principal bundle over $\Xi$ with structure group $A(n)$ = affine group that is bundle-isomorphic to the affine frame bundle of $\Xi$. We do not use the affine frame bundle of $\Xi$ as $P$, because the affine frame bundle has a fixed solder form, which section 2.6.2 shows is not appropriate for EC. See appendix C for a discussion of global equivalence of fibre bundles. Let $X$ = flat affine vector space on which $A(n)$ acts effectively. Let $x^i$ be affine coordinates on $X$.

A vector field in $L(A(n)) \cdot X$ canonically (that is, without any arbitrary choices) defines a homogeneous linear component, derived as the limit of the action of the vector field on large spheres centered on any point in $X$. However, an element of $L(A(n)) \cdot X$ does not define a unique translational part unless we choose a zero section of $X$. This reflects the facts that

- The subgroup $A_0(n)$ of translations with zero homogeneous part is canonically defined, whereas the subgroup $GL(n)$ of homogeneous transformations without translation of the origin is not canonically defined.

- The canonical exact sequence of groups

(1) $$0 \rightarrow A_0(n) \xrightarrow{\alpha} A(n) \xrightarrow{\beta} GL(n) \rightarrow 0$$

is not canonically split; that is, defining a map $\gamma : GL(n) \rightarrow A(n)$ such that $\beta \cdot \gamma =$ identity ($\gamma$ followed by $\beta$) requires an arbitrary choice – the choice of a preferred origin point in an affine representation space for $A(n)$ to serve as the fixed point of $GL(n)$.

### 2.2. Choose an Affine Connection

Choose an affine connection 1–form $\omega$ on $P(\Xi, A(n))$ and denote its connection

---

[2]    Reference (Petti 1976) clearly states that the zero section violates translational equivariance / gauge symmetry. The statement in section 2.7.3 that Einstein Cartan theory is a gauge theory of the Poincaré group should have been accompanied by the appropriate qualification to that effect.

[3]    $G$ acts effectively on $X$ if and only if only the identity element of $G$ maps onto the identity transformation of $X$. If $G$ is connected, it acts effectively on $X$ if and only if only the zero element of $L(G)$ maps onto the zero vectorfield on $X$.

coefficients by $\Gamma_\mu{}^i(\xi, x)$.

An affine connection uniquely defines a homogeneous linear connection, but not a translational part unless we choose a zero section of the associated affine vector bundle. Up to this point, the construction preserves full affine equivariance / gauge invariance.

### 2.3.  *Choose a 'Zero Section'*

Choose a 'zero section' $s : \Xi \rightarrow B(\Xi, A(n), P, X)$, which defines a preferred point in each fibre that is the fixed point of the action of $GL(n)$.  Choose the fibre coordinates so $x^i(s(\xi)) = 0$ in the fibre over each $\xi \in \Xi$ .

The choice of $s$ violates affine equivariance / gauge invariance. $s$ splits $\omega$ into translational and homogeneous linear components. In coordinates $(\xi^\mu, x^i)$ on $B(\Xi, A(n), P, X)$, the connection coefficients have the form

(2)                        $\Gamma_\mu{}^i(\xi, x) = K_\mu{}^i(\xi) + B_{\mu\,j}{}^i(\xi)\, x^j$

where the translational connection coefficients are $K_\mu{}^i(\xi) = \Gamma_\mu{}^i(\xi, s(\xi))$ and the linear connection coefficients are $B_{\mu\,j}{}^i(\xi)$. The curvature tensor of $\omega$ has the form

(3)                        $R_{\mu\nu}{}^i(\xi, x) = T_{\mu\nu}{}^i(\xi) + R_{\mu\nu\,j}{}^i(\xi)\, x^j$

where $T_{\mu\nu}{}^i(\xi)$ is the translational curvature (affine torsion) and $R_{\mu\nu\,j}{}^i(\xi)$ is the (homogeneous) linear curvature.

An affine gauge transformation $(r,\ a)$ (where $r$ is a homogeneous linear transformation, and $a$ is a translation vector) defines new coordinates on the affine tangent bundle by $x'^i = (r^{-1})^i{}_j\, x^j - a^i$. The affine connection coefficients in the new coordinates are

(4)                $K'_\mu{}^i(\xi) = r^{-1}(\xi)^i{}_k\, (B_{\mu h}{}^k(\xi)\, a^h(\xi) + K_\mu{}^k(\xi) + \partial_\mu\, a^k(\xi)\, )$

(5)                $B'_{\mu\,j}{}^i(\xi) = r^{-1}(\xi)^i{}_k\, (B_{\mu h}{}^k(\xi)\, r^h{}_j(\xi) + \partial_\mu\, r^k{}_j(\xi)\, )$

In particular, if $r = identity$ then

(6)                $K'_\mu{}^i(\xi) = K_\mu{}^i(\xi) + B_{\mu\,j}{}^i(\xi)\, a^j$                        $B'_{\mu\,j}{}^i(\xi) = B_{\mu\,j}{}^i(\xi)$ .

The $K_\mu{}^i(\xi)$ define a linear homomorphism from $T_\xi\Xi$ to $L(A_0) \cdot X$, or equivalently to the fibre $T_{s(\xi)}B(\Xi, A(n), P, X)_\xi$.

Equation (6) captures the main complication that arises from including translational symmetries in gauge theories: a zero section $s$ of the associated affine bundle enables us to define the translational part of an affine connection $\omega$ as $K_\mu{}^i(\xi) := \omega_\mu{}^i(\xi, s(\xi))$; but $s$ breaks full translational equivariance / translational gauge invariance. Conversely, specifying $K_\mu{}^i(\xi)$ does not uniquely determine $s(\xi)$ if, for some vector $\delta s^j(\xi)$, $B_{\mu\,j}{}^i(\xi)\,\delta s^j(\xi) = 0$. Specifying $K_\mu{}^i(\xi)$ avoids compromising translational equivariance only if all linear connection coefficients $B_{\mu\,j}{}^i(\xi)$ are zero.

### 2.4.  *Require Translational Connection Coefficients K to define an Invertible Mapping*

For a general affine bundle, appendix B outlines how to define an associated solder form $\Theta$ in terms of the zero section $s$ and the translational part $K_\mu{}^i$ of the connection $\omega$, even if the affine bundle is not globally equivalent to the tangent bundle of $\Xi$. We require the associated solder form to be invertible, which is equivalent to $K_\mu{}^i$ being invertible. Invertibility of $K_\mu{}^i(\xi)$ enables us to

•    Identify the tangent space $T_\xi\Xi$ with $L(A_0) \cdot X_\xi$ (where $X_\xi$ is the fibre of $B(\Xi, G, P, X)$ over $\xi \in \Xi$)

•    Define a local solder form $\theta$ in terms of $K$.

•    Treat the inverse of the translational connection coefficients $K_i{}^\mu(\xi)$ as a linear frame field.

In order for the connection and zero section to define a solder form globally, hence to make $P(\Xi, A(n))$ the affine frame bundle of $\Xi$, the connection must satisfy some global conditions. See appendix C for a brief discussion of global equivalence of bundles.

For a general affine bundle, both the affine connection $\omega$ and the zero section $s$ determine whether $K_\mu{}^i$ is invertible. Here is an example in two dimensions where the choice of $s$ can make $K_\mu{}^i$ non-invertible. Let $K_\mu{}^i = \delta_\mu{}^i$ ,

$$B_1{}^j = \begin{vmatrix} 0 & -b_1 \\ b_1 & 0 \end{vmatrix}, B_2{}^j = \begin{vmatrix} 0 & -b_2 \\ b_2 & 0 \end{vmatrix}. \text{ If } s^i = \begin{vmatrix} a^1 \\ a^2 \end{vmatrix} \text{ then } K'_\mu{}^i(s) = \begin{vmatrix} 1 - b_1 \, a^2 & - b_2 \, a^2 \\ b_1 \, a^1 & 1 + b_2 \, a^1 \end{vmatrix}.$$

If $a^1 \, b_2 - a^2 \, b_1 = -1$, then $K'_\mu{}^i$ is not invertible.

### 2.5. Define a Fibre Metric

Let $g_{ij}(\xi)$ be a non-degenerate symmetric bilinear form on the tangents of each fibre of $B(\Xi, A(n), P, X)$. $K_\mu{}^i$ pulls back $g_{ij}$ to a (pseudo–)metric $g_{\mu\nu}(\xi) = K_\mu{}^i(\xi) \, g_{ij}(\xi) \, K_\nu{}^j(\xi)$ on $\Xi$. Require that $\omega$ preserves $g_{ij}(\xi)$, which occurs if and only if $B_\mu{}^k{}_j \, g_{ki}$ is antisymmetric in $i$ and $j$.

### 2.6. Construct Einstein-Cartan Theory

*2.6.1. Construction.* As the master theory of classical physics, GR has one known flaw: it cannot describe spin-orbit coupling properly because of the symmetry of the stress tensor. In 1922 E. Cartan conjectured that general relativity should be extended to include affine torsion to solve this problem. The resulting EC theory appeared to require an independent assumption beyond GR, and effects are too small to measure at this time (Kerlick 1975); so the theory was long considered a speculative extension of GR.

Petti proved that GR plus matter with spin imply EC theory with no further assumptions (Petti 1986, a factor of ½ corrected in Petti 2001). The construction yields nonlinear (in angular momentum) expressions for the translational holonomy of an isolated rotating black hole, which is an integral surrogate for affine torsion. The final stage of passing to the fluid continuum limit yields the spin-torsion equations of EC. [4]

Using Lagrangian $L = \frac{1}{2} R - 8 \pi K_{grav} L_{matter}$, vary the action $A = \int \sqrt{g} \, d^4\xi \, L$ with respect to the linear and translational connection coefficients to obtain the field equations of EC.

$$\delta A / \delta B_\mu{}^{ij} = G_i{}^\alpha - 8 \pi G \, P_i{}^\alpha = 0 \tag{7}$$

$$\delta A / \delta K_\mu{}^i = S_{ij}{}^\alpha = 8 \pi G \, J_{ij}{}^\alpha = 0, \tag{8}$$

where $G_i{}^\alpha$ is the Einstein curvature tensor, $S_{ij}{}^\alpha = T_{ij}{}^\alpha + g_i{}^\alpha \, T_{j\gamma}{}^\gamma - g_j{}^\alpha \, T_{i\gamma}{}^\gamma$ is the modified torsion tensor, $T_{ij}{}^\alpha$ is the affine torsion tensor (equivalently, the translational curvature), $P_i{}^\alpha$ is the momentum tensor, and $J_{ij}{}^\alpha$ is the spin tensor of matter and radiation.

EC provides strong motivation for including translational symmetries in spacetime theories. The equations of EC are simpler and make more sense as a gauge theory if we distinguish base space tensors and fibre tensors that are related via a frame field (Petti 1976).

---

[4]    Adamowicz showed that GR plus a linearized classical model of matter with spin yields the same linearized equations for the time-time and space-space components of the metric as linearized EC (Adamowicz 1975). Adamowicz does not mention the time-space components of the metric, the spin-torsion field equation, spin-orbit coupling and the non-symmetric momentum tensor, the geometry of torsion, or quantum mechanical spin. He says, "It is possible a priori to solve this problem [of dust with intrinsic angular momentum] exactly in the formalism of GR but in the general situation we have no practical approach because of mathematical difficulties." Adamowicz's conclusion is at best incomplete: it is not possible to solve the full problem exactly in GR, including spin-orbit coupling, without adopting the larger framework of EC theory.

### 2.6.2. *Does Einstein–Cartan Theory Use an Affine Frame Bundle?* A key question is whether EC uses

- an affine frame bundle that is endowed with a fixed solder form θ; or
- an affine bundle, and an associated solder form Θ defined in terms of an affine connection $\omega$ and zero section $s$.

Let us examine both alternatives, being careful about how we identify tangents to the base manifold $\Xi$ and tangents to the associated affine vector space $X$. To construct the Lagrangian, both approaches begin with the homogeneous curvature tensor $R_{\mu\nu j}{}^i$, as defined in equation (B.6) in appendix B. When defining the scalar curvature, the two approaches differ on how to identify base space and fibre tangents.

- If EC uses an affine frame bundle, it has a solder form $\theta$, which defines a metric on $\Xi$ by $g_{\mu\nu} = \theta_\mu{}^i\, g_{ij}\, \theta_\nu{}^j$. Then the scalar curvature is $R = R_{\mu\nu j}{}^i\, \theta_i{}^\mu\, \theta_m{}^\mu\, g^{mn}\, \theta_n{}^\nu$. Variation of $R$ by $B_{\mu j}{}^i$ yields the spin-torsion field equations. However, $g_{ij}$ and $\theta$ are fixed and neither can be varied.

- If EC uses an affine bundle not endowed with a fixed solder form, then we choose a zero section $s$ and a connection $\omega$, use them to define $K_\mu{}^i$, require that $K$ be invertible, and define a metric on $\Xi$ by $g_{\mu\nu} = K_\mu{}^i\, g_{ij}\, K_\nu{}^j$. Then the scalar curvature is $R = R_{\mu\nu j}{}^i\, K_i{}^\mu\, K_m{}^\mu\, g^{mn}\, K_n{}^\nu$. Variations of $R$ by $K_\mu{}^i$ and $B_{\mu j}{}^i$ yield the curvature-momentum and spin-torsion EC field equations respectively, as in equations (7) and (8) in section 2.6.

Sometimes the 1-1 tensor field that pulls the fibre metric $g_{ij}$ back to $\Xi$ is referred to as a frame field, as if it is independent of the solder form. However, the solder form defines an isomorphism between $T_\xi\Xi$ and the translational vector fields $L(A_0) \cdot X$, so is used to pull back $g_{ij}$ to $g_{\mu\nu}$. Using an independent frame field for this purpose is unnatural.

This formulation makes clear that EC must use an affine bundle without a fixed solder form. Lifting the associated solder form to a tensorial form on the principal bundle defines the solder form. The intuitive rationale for a dynamical solder form in the basic structure of EC is that EC is about piecing together local high-symmetry versions of spacetime (the fibres) to form a spacetime. The affine connection $\omega$ and zero section $s$, through the variational process, determine how the flat local fibres fit together to construct the base manifold.

### 2.7. *Symplectic Structure*

We construct a symplectic structure from the fields $K$ and $s$ at two levels of generality. Throughout this paper,

- The "General Case" means that $G$ is a Lie group with closed subgroup $G_1$, and $G$ acts effectively on the left coset manifold $X = G/G_1$. $X$ need not have a linear or affine structure.

- The "Affine Case" means that $G$ is the affine group $A(n)$, $G_1$ is $GL(n)$, and $X$ is an $n$-dimensional affine vector space on which $G$ is represented.

### 2.7.1. *General Case.* Over the bundle $B(\Xi, A(n), P, T^*{}_sX)$, define a complex unitary line bundle $B(B(\Xi, A(n), P, T^*{}_sX), U(1), \mathbf{C})$. Here, $T^*{}_sX$ denotes the cotangent space to $X$ at the point $s(\xi) \in X$. Define a connection form on this bundle

$$(9) \qquad\qquad \varphi 1 := \mathrm{i}\, p_i\, K_\mu{}^i\, \mathrm{d}\xi^\mu, \qquad \text{where } p_i \in T^*{}_{s(\xi)}X_\xi$$

so that $\mathrm{i}\, p_i\, K_\mu{}^i$ are its connection coefficients. The curvature of the connection is

$$(10) \qquad \varphi 2 := \mathrm{d}\ \varphi 1 = \mathrm{i}\ \mathrm{d}p_i\, K_\mu{}^i\, \mathrm{d}\xi^\mu \boxed{+\ \mathrm{i}\ p_i\ {\textstyle\frac12}\ (K_{\nu^\cdot,\mu}{}^i - K_{\mu^\cdot,\nu}{}^i)\, \mathrm{d}\xi_2^\mu\, \mathrm{d}\xi^\nu.} \quad \text{\scriptsize Missing in publication}$$

φ1 is the symplectic 1-form and φ2 is the symplectic 2-form. The Bianchi identify is dφ2 ≡ 0, which is the condition that the symplectic 2-form is closed. Because φ1 depends on $K$, the symplectic structure depends on the connection form and the zero section $s$.

*2.7.2. Affine Case* Everything is the same as the general case, except that we start with a complex unitary line bundle $B(T^*\Xi, U(1), \mathbf{C})$ over the bundle cotangent bundle $T^*\Xi$, so that $p_i \in T^*_\xi \Xi$.

In the path integral formulation of quantum mechanics, momentum is the rate of change of complex phase with position. This change of phase defines a parallel translation in a complex line bundle over phase space.[5] Because of gauge invariance with respect to complex phase, physical results depend only on the symplectic 2–form and never on the symplectic 1–form.

### 2.8. Roles of the Zero Section s and the Connection Coefficients K

The zero section $s$ breaks the translational equivariance of any construction in which it is used. It is used in the following operations in the general case and in the affine case.

- $s$ reduces the principal $G$-bundle $P$ to a bundle with group $G_1$.

- Given an affine connection $\omega$ with connection coefficients $\Gamma_\mu{}^i$, $s$ defines $K_\mu{}^i$ by $K_\mu{}^i(\xi) := \Gamma_\mu{}^i(\xi, s(\xi))$.

  – $K_\mu{}^i$ defines the associated solder form $\Theta$.
  – If $K_\mu{}^i$ is invertible then the lift of $\Theta$ can serve as a solder form $\theta$, and $K_\mu{}^i$ functions as an (inverse) frame field. This still does not guarantee that the affine bundle is the affine tangent bundle. That requires that $K_\mu{}^i$ can be the identity transformation in every coordinate patch.
  – $K_\mu{}^i$ defines the symplectic 1-form and 2-form.

## 3. Cartan Connections

The usual mathematical structure for including translational symmetries in connections on fibre bundles is a "Cartan Connection." The Cartan connection serves as the basis for EC.

### 3.1. Definition of Cartan Connections

The construction in sections 2.1–2.5 is equivalent to defining a Cartan connection in differential geometry, which is defined as follows (Kobayashi 1972, pages 127-128).

Let $\Xi$ be an *n*-manifold, $G$ a Lie group (e.g. the affine group $A(n)$), and $G_1$ a closed subgroup of $G$ (e.g. the homogenous linear group $GL(n)$) so that $G/G_1$ has dimension $n$. Let $P$ be a principal bundle over $\Xi$ with group $G_1$.

---

[5] To see this, perform this experiment: Starting at $(q,p)$, parallel translate $z$=1 to the point $(q+\delta q, p)$ where $z$=1–i$p\delta q$; to $(q+\delta q, p+\delta p)$ where $z$ is unchanged; to $(q, p+\delta p)$, where $z$=1–i$p\delta q$+ i$(p+\delta p)\delta q$=1+i$\delta p\delta q$; and back to $(q,p)$ where still $z$=1+i$\delta p\delta q$. So $\delta z$=+i$\delta p\delta q$, which illustrates that the curvature form is +id$p$^d$q$.

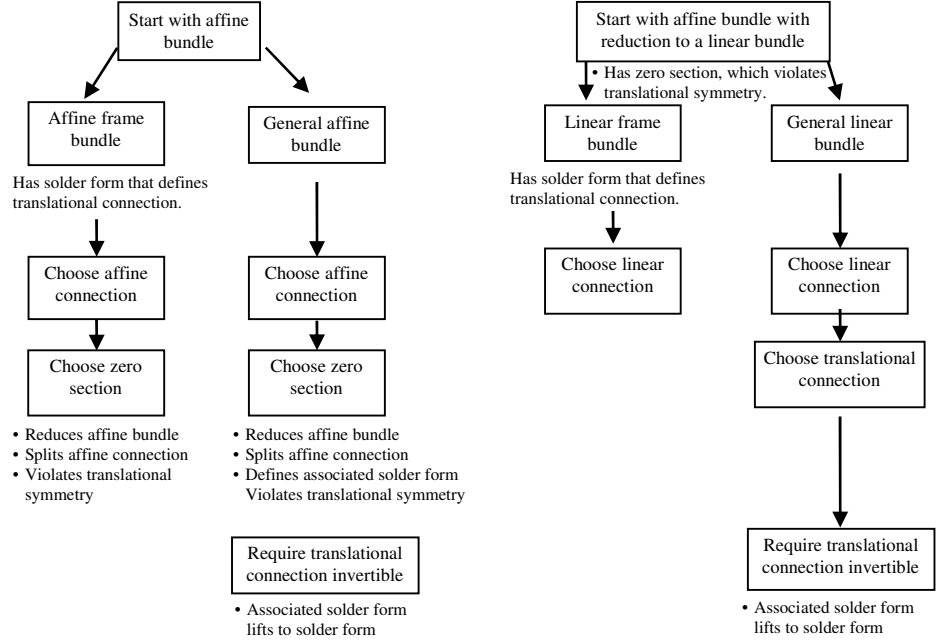

**Figure 1:** Four ways to construct a Cartan Connection

Define a Cartan connection as follows. Let $\omega$ be a 1–form on $P$ that takes values on the Lie algebra $L(G)$, such that

- $\omega$ is a vertical form; that is, for every $v \in L(G_1)$, $\omega(v^*) = v$, where $v^*$ is the vertical vector field induced on $P$ by $v$ and the action of $G_1$ on $P$.

- $\omega$ is equivariant with respect to the action of $G_1$; that is, $(R_a)^* \omega = ad(a^{-1}) \omega$, where $a \in G_1$, $R_a$ denotes the right action of $a$ on P, and $ad$ is the adjoint representation of $G$ on $L(G)$.

- At each point $p$ of $P$, $\omega$ is an isomorphism of the tangent space $T_p P$ and $L(G)$; that is, for all $y \neq 0$ tangent to $P$, $\omega(y) \neq 0$.

The horizontal and equivariance properties are the same as in the definition of ordinary connections, except that the Cartan connection form $\omega$ takes values in the larger Lie algebra $L(G)$.

The isomorphism condition causes the translational connection coefficients to define an invertible mapping between tangents to $P$ and $L(G)$, so $L(G)$ serves as a high-symmetry local model for the affine frames of $\Xi$.

### 3.2. Multiple Ways to Define Cartan Connections

We can define a Cartan connection either starting with an affine bundle or a linear bundle. The bundle can be the affine or linear frame bundle or a general affine or linear bundle. In each case, we end up with a zero section, a solder form, and a linear connection. The diagram below captures four ways of defining a Cartan connection when the translational part of the connection determines the solder form.

Section 2 in the main body of this paper traces through the second path from the left. This path is the most general case and it preserves full affine symmetry as long as possible, until we choose a zero section. This path also defines the solder form in terms of the connection form and the zero section, through the associated solder form. If the connection form varies, then the solder form varies.

## 4. Critique of Other Approaches

Most of the confusion about translational symmetries in the literature is due to two factors.

- Many papers discuss gauge symmetry in terms of "how fields transform," not in terms of bundle structure. While a description with equations and variables is usually superior for computation, a description with sets and mappings (and bundles for differential geometry) is superior for treating convergence and coordinate-invariant group symmetries, and for compressing enormous computational complexity. Choosing a vector space basis or manifold coordinate system recovers equations and variables from abstract descriptions.

- Some authors assume that the affine bundle $P$ can be reduced to a linear sub-bundle without choosing a zero section of the affine tangent bundle $B$. Therefore these authors construct something like a Cartan connection, but they seem to believe that they have preserved the translational gauge symmetry.

We select two works for discussion: the lengthy survey of metric affine gravity by Hehl *et al* and a recent paper by Tresguerres that develops the approach of Lord and Goswami.

### 4.1. The Approach of Hehl and Ne'eman

Ne'eman had several objections to using the zero section $s$ and the translational connection coefficients to define a frame field (Ne'eman 1979). (1) The Einstein Lagrangian is not invariant under translational gauge transformations unless the equations of motion are employed. (2) The symmetry group contains only Poincaré translations of the Minkowski fibre $X$, and not finite translations of the spacetime manifold $\Xi$. (3) The linear connection coefficients have zero variation with respect to translations. (4) The approach fails when spinning matter is present.

The author believes that these objections reflect no more than an aesthetic preference that full translational gauge symmetry be preserved. (1) The fact that the frame field identifies spacetime tangents and fibre Poincare translations is adequate to include spacetime translations in the theory. (2) There is no reason why the Einstein Lagrangian $R$ must be invariant under anholonomic Poincaré gauge translations. (3) There is no reason why the linear connection coefficients cannot have zero variation with respect to Poincaré translations / frame fields. (4) We can represent the spin covering group of the Poincaré group on spinor fields defined on the fibre space $X$.

The extensive survey of metric affine gravity by Hehl et al. (Hehl et al. 1995), of which Ne'eman is an author, develops this point of view. We believe this approach has the following defects.

- The splitting of the connection into translational and linear parts (in their equation (3.2.2)) requires a zero section $s$ that breaks the translational equivariance / gauge invariance. Hehl et al. seem to believe they have defined a theory that preserves translational symmetry.

- The zero section $s$ appears in their theory as a "zero form $\xi$" (in their equation (3.2.10)). However, they do not identify its role in splitting the affine group, in defining the solder form, or in defining the origin of linear coordinates on affine fibres. This field seems to be an arbitrary gauge degree of freedom with no other role in the theory.

- The translational curvature consists of the torsion and an additional term (in their equation (3.2.13)). It appears that this extra term arises from permitting the translational part of the connection $K$ to differ from the solder form $\theta$, which we explain in appendix B seems unnatural.[6]

*4.2. Tresguerres' Composite Bundles*

Tresguerres objected to the approach based on Cartan connections that is outlined in section 2 (Tresguerres 2002).[7]

> "Indeed, neither the geometric meaning of $[K_i{}^a]$ as a vierbein nor the universal coupling of gravitation to the remaining forces receives an explanation in the standard bundle approach. We know such problems to be absent from the frame bundle treatment of gravity, where tetrads are sections rather than connections. The price one has to pay when adhering to this view is that one must accept gravitational potential of two kinds, the tetrad potentials (sections) being different in mathematical nature from those of other interactions (connections), such tetrads having nothing to do with the translations in the Poincaré group."
> (page 5)

In an apparent attempt to preserve translational equivariance / gauge symmetry, Tresguerres introduces a "composite bundle" consisting of a tower of bundles $P(\Sigma, G_1) \rightarrow \Sigma(\Xi, G/G_1) \rightarrow \Xi$, where $\Xi$[8] is spacetime, $G$ is the Poincaré group, $G_1$ is the Lorentz group.

> "Roughly speaking, our leading idea is that of attaching to each point of the base space $\Xi$ [M in Tresguerres' notation] a fibre with the bundle structure $G(G/G_1, G_1)$."
> (page 5)

The material in appendix B shows how the zero section $s$ relates the translational connection coefficients $K$ and the solder form $\theta$, so that $K$ can be used as linear frame fields so long as the zero section $s$ is fixed. It also shows that Tresguerres' tower of bundles violates the translational symmetry because it implicitly chooses a zero section $s : \Xi \rightarrow B(\Xi, G, P, X)$.

## 5. Conclusion

The author recommends that gravitational physicists adopt the standard differential geometric construction of Cartan connections to include translational symmetries in spacetime physics. The identification of infinitesimal spacetime translations with anholonomic Poincaré translations is adequate to establish the theory as a theory of translational symmetry. This method yields variational equations and field equations that express EC as an affine theory, not a metric theory, of gravitation with spin. There is no need to employ actual translations of spacetime to represent translational symmetry. Translational gauge symmetry of the Lagrangian is violated. This viewpoint is similar to that used by physicists for crystals with defects, where fibres represent idealized high-symmetry neighborhoods (perfect crystal grains), discrete holonomy and continuous curvature represent line defects (dislocations, disclinations, and dispirations).

The approach outlined here uses the key concept of the zero section $s$ to clarify the relationship between affine and linear connections. It derives several basic concepts from the zero section $s$ and the affine connection $\omega$, namely the reduction of the affine bundle to a linear bundle, the splitting of the affine connection into translational and linear parts, the definition of the solder form, and definitions of the symplectic 1-form and the symplectic 2-form. This approach generalizes these concepts to a more general setting than affine and linear bundles. It also rules out defining a solder form or symplectic forms independently of the zero section and the affine the connection, unless there is a compelling reason to do so.

---

[6]   Both the solder form and the translational connection coefficients have the main purpose of identifying base space vectors and fibre vectors.

[7]   Tresguerres first points out the deficiencies of Lord's point of view (Lord 1985, 1986, 1986a, 1988) in which (1) the base manifold is G / H $\approx$ R$^4$ , and (2) gauge transformations can map a fibre onto a different fibre.

[8]   Tresguerres denotes spacetime by M with coordinates x$^i$. The present author denotes spacetime by $\Xi$ with coordinates $\xi^\mu$, and employs Roman symbols for the fibre space X with coordinates x$^i$.

**Acknowledgements**

I would like to thank David Bleecker for discussions of the differential geometry of translational symmetries in fibre bundles. I would also like to thank one of the reviewers for suggesting clarifications of the conditions for invertibility of the translational connection coefficients and other matters.

## Appendix A. Connections and Equivariance

### A.1 Definition of Equivariance and Gauge Invariance

We begin with some definitions of connections and equivariance from chapter 2 of Kobayashi and Nomizu (1963) and chapter 5 of Bishop and Crittenden (1964).

*Definition*: A *connection* on principal bundle $P(\Xi, G)$ is defined as a subspace $H_p$ of $T_p(P)$ for each $p \in P$ such that

(a)  $H_p$ depends differentiably on $p$.

(b)  $T_p(P) = V_p \oplus H_p$ (direct sum), where $V_p$ is the space of tangent vectors in $T_p(P)$ induced by the action of $G$ on $P$. ($V_p$ is called the space of vertical vectors tangent to $P$ at $p$, and $H_p$ is called the space of horizontal vectors tangent to $P$ at $p$.)

(c)  The mapping $p \rightarrow H_p$ is invariant under the action of $G$ on $P$. That is, for every $p \in P$ and $g \in G$, $H_{pg} = (\mathrm{d}R_g) * H_p$ , where $\mathrm{d}R_g$ is the transformation of $P$ induced by $g \in G$, where $R_g\, p = p \cdot g$.

*Definition*: The *1-form of a connection* is a Lie algebra-valued 1-form $\omega$ on $P$ defined by (1) projecting $T_p(P)$ onto $V_p$, and (2) mapping each element of $V_p$ to the element of $L(G)$ that induces that vector in $T_p(P)$.

*Definition*: A $L(G)$-valued 1-form $\omega$ on a principal bundle P is *equivariant* if $\omega\, \mathrm{d}R_g = ad(g^{-1})\, \omega$.

The 1-form of a connection is equivariant. Condition (b) above and the equivariance property of the connection 1-form express that the connection and its 1-form are invariant under the action of $G$ on $P$. This generalizes the facts that computations using sections of an associated linear (affine) vector bundle can be expressed using any differentiable linear (affine) bases in the fibres. (An affine basis is a choice of origin point and a linear vector space basis in each fibre.)

### A.2 Local Coordinate Expressions for the 1-form of a Connection

Let $g^A{}_B$ be a matrix representation of $G$. Choose a section $z : \Xi \rightarrow P$ and define coordinates $g^A{}_B$ on $V_p$ so that $g^A{}_B(z(\xi)) \equiv \delta^A{}_B$; then the action of $G$ on $P$ induces coordinates $g^A{}_B$ on the vertical fibres $V_p$ of $P$. The section $z$ has no coordinate-invariant significance. The 1-form of the connection at point $(\xi^\mu, g^A{}_B) \in P$ can be written in local coordinates as

$$(A.1) \qquad \omega_\mu{}^A{}_B = (g^{-1})^A{}_C\, (\mathrm{d}g^C{}_B \; + \mathrm{d}\xi^\mu\, \Gamma_\mu{}^C{}_D\, g^D{}_B) \,,$$

where $\Gamma_\mu{}^A{}_B$ are the connection coefficients. In bundle coordinates, invariance under the symmetries of $G$ means we can change the bundle coordinates by choosing a different section $z' \equiv z \cdot \zeta$, where $\zeta : \Xi \rightarrow G$. In the new coordinates,

$$(A.2) \qquad \Gamma'_\mu{}^A{}_B = (\zeta^{-1})^A{}_C\, (\Gamma_\mu{}^C{}_D\, \zeta^D{}_B + \partial_\mu \zeta^C{}_B) \,.$$

### A.3  Homogeneous Linear Case

Assume the associated fibre $X$ is a linear vector space with linear coordinates $x^i$, and that $G$ above acts on $X$ with matrix representation $g^i{}_j$.

*Definition*: If $P$ is the linear frame bundle of $\Xi$, the *solder form* (or canonical 1-form) $\theta : TP \to R^n$ is

(A.3)                                                       $\theta = p^{-1} \cdot \mathrm{d}\pi$ ,

where $p \in P$ defines the mapping $p : R^n \to T_{\pi(p)}\Xi$.

The solder form is equivalent to an isomorphism between each tangent space $T_\xi\Xi$ and $X$. (See Example 5.2 in Chapter 2 of Kobayashi and Nomizu (1963).)

### A.4  (Inhomogeneous) Affine Case

Assume the associated fibre $X$ is an affine vector space with affine coordinates $x^i$, and that $G$ acts as affine transformations on $X$. Let $(h^i{}_j, t^i)$ be homogeneous and translational components of the matrix representation; that is, the element of $G$ with coordinates $(h^i{}_j, t^i)$ acts on $x \in X$ by $x'^i = h^i{}_j x^j + t^i$. The distinction of homogenous and translational coordinates on $G$ is coordinate-dependent. If we choose a section $z : \Xi \to P$, then $(h^i{}_j, t^i)$ defines coordinates on the vertical fibres $V_p$ of $P$, and the connection form can be written in terms of $K_\mu{}^i$ and $B_\mu{}^i{}_j$ as in equation (2). At point $(\xi^\mu, h^i{}_j, t^i) \in P$, the 1-form of the connection can be written in local coordinates in terms of the connection coefficients $\Gamma_\mu{}^i{}_j$ as

(A.4)                          $\omega^i{}_j = (h^{-1})^i{}_k \, (\mathrm{d}r^k{}_j + \mathrm{d}\xi^\mu \, \Gamma_\mu{}^k{}_m \, h^m{}_j)$

(A.5)                          $\omega^i = (h^{-1})^i{}_k \, (\mathrm{d}t^k + \mathrm{d}\xi^\mu \, (\Gamma_\mu{}^k{}_m \, t^m + \Gamma_\mu{}^k) \, )$

We can change the bundle coordinates by choosing a different section $z' \equiv z \cdot (\eta, \tau)$, where $(\eta, \tau) : \Xi \to G$ are the homogeneous and translational parts of the mapping. In the new coordinates,

(A.6)                          $\Gamma'_\mu{}^i{}_j = (\eta^{-1})^i{}_k \, (\Gamma_\mu{}^k{}_m \, \eta^m{}_j + \partial_\mu \eta^k{}_j)$ .

(A.7)                          $\Gamma'_\mu{}^i = (\eta^{-1})^i{}_k \, (\Gamma_\mu{}^k{}_m \, \tau^m + \Gamma_\mu{}^k + \partial_\mu \tau^k)$ .

The origin in each fibre depends on the choice of coordinate system. If we want a zero vector and scalar multiplication in each fibre, we must violate translational equivariance / gauge invariance by choosing an origin in each fibre.

*Definition*: If $P$ is the affine frame bundle of $\Xi$, the *affine solder form* $\theta : TP \to R^n$ is

(A.8)                                                       $\theta = p^{-1} \cdot \mathrm{d}\pi$ ,

where $p \in P$ defines the mapping $p : R^n \to T_{\pi(p)}\Xi$.

The affine solder form is equivalent to an isomorphism between each tangent space $T_\xi\Xi$ and the space of vector fields $L(A_0) \cdot X$ generated on $X$ by $L(A_0(n))$, where $A_0(n)$ is the group of translations with vanishing homogeneous linear transformations. When $X$ is a homogeneous linear space, $T_\xi\Xi$ is isomorphic to both $X$ and $L(A_0) \cdot X$. When $X$ is an affine vector space, we see that the correct generalization for linear and affine cases is that $T_\xi\Xi$ is isomorphic as a linear space to $L(A_0) \cdot X$, and not to $X$.

## Appendix B.  Connections, Zero Sections and Associated Solder Forms

A description of connections in terms of associated bundles helps develop intuition.

### B.1  General Case

Let $G$ be a Lie group with closed subgroup $G_1$, so that $G$ acts effectively on the left coset manifold $X = G/G_1$.

Choose a connection form $\omega$ on $P$ and a smooth section $s : \Xi \to B(\Xi, G, X)$, called the "zero section." The sections $s$ are in one-to-one correspondence with reductions of the principal bundle $P$ to a principal bundle with structure group $G_1$. (See proposition 5.6 and the following remark in Chapter 1 of Kobayashi and Nomizu (1963).) We generalize the definition of a solder form.[9]

*Definition*: The associated solder form $\Theta : T_\xi X \to T_{s(\xi)} B(\Xi, G, X)_\xi$ is given by

(B.1) $\qquad\qquad \Theta_\mu{}^i(\xi) := -\nabla_\mu\, s^i(\xi) = -\,\partial_\mu\, s^i(\xi) + \Gamma_\mu{}^i(\xi, s(\xi)) \ .$

In fibre coordinates for which $x^i(s(\xi)) = 0$ for all $\xi$, this becomes

(B.2) $\qquad\qquad\qquad\qquad \Theta_\mu{}^i(\xi) = K_\mu{}^i(\xi) \ ,$

where $K_\mu{}^i(\xi) := \Gamma_\mu{}^i(\xi, s(\xi))$ . $\Theta$ need not be an isomorphism because the dimensions of $\Xi$ and $X$ are not constrained to be equal.

### B.2  Affine Case

Let $G$ be the affine group $A(m)$, $G_1$ be $GL(m)$, and $X$ be an affine vector space of dimension $m$. ($m$ need not equal $n = dim(\Xi)$). Perform all the constructions in the previous section. If the coordinates $x^i$ on $X_\xi$ have value zero at the point $s(\xi)$, then the connection can be written as

(B.3) $\qquad\qquad\qquad \Gamma_\mu{}^i(\xi, x) = K_\mu{}^i(\xi) + B_{\mu\,j}{}^i(\xi)\, x^j$

The curvature tensor of the connection can be written as

(B.4) $\qquad\qquad\qquad R_{\mu\,\nu}{}^i(\xi, x) = T_{\mu\,\nu}{}^i(\xi) + R_{\mu\,\nu\,j}{}^i(\xi)\, x^j \ .$

where the affine torsion $T_{\mu\,\nu}{}^i$ and the homogenous linear curvature $R_{\mu\nu}{}^i$ are given by

(B.5) $\qquad T_{\mu\nu}{}^i(\xi) = \partial_\mu K_\nu{}^i(\xi) - \partial_\nu K_\mu{}^i(\xi) + B_{\mu k}{}^i(\xi)\, K_\nu{}^k(\xi) - B_{\nu k}{}^i(\xi)\, K_\mu{}^k(\xi)$

(B.6) $\qquad R_{\mu\nu\,j}{}^i(\xi) = \partial_\mu B_{\nu\,j}{}^i(\xi) - \partial_\nu B_{\mu\,j}{}^i(\xi) + B_{\mu k}{}^i(\xi)\, B_{\nu\,j}{}^k(\xi) - B_{\nu k}{}^i(\xi)\, \Gamma_{\mu\,j}{}^k(\xi) \ .$

The associated solder form $\Theta$ is defined for any affine bundle (not just the affine tangent bundle) that has a zero section $s$ and an affine connection $\omega$.

On an affine bundle which comes equipped with a solder form $\theta$, we can define an affine connection whose translational part is independent of the solder form. However, we then have two 2-index tensors that require variational principles or at least field equations to specify them. The construction of the associated solder form shows how the associated solder form (and its lift to the principal bundle) arises from the connection and the zero section as the natural way to identify tangents to the base space and tangents to the fibre.

A similar relation in section 2.7 defines the symplectic 1-form and 2-form in terms of $s$ and $\omega$.

---

[9]   The conventional definition of the solder form is a tensorial 1-form $\theta : P(\Xi, GL(n)) \to R^n$, where $n = dim(\Xi)$. (See "solder form" in Bishop and Crittenden (1964) or "canonical form" in Kobayashi and Nomizu (1963).)

### Appendix C.  Global Equivalence of Fibre Bundles

Our main purpose is to explore the local field theory over a simply connected spacetime. However, we make some basic observations about global bundle structure. The key global issues are (1) whether bundles are equivalent, and (2) what distinguishes the tangent bundle of a spacetime manifold $\Xi$. See Steenrod (1951) for an introduction to fibre bundles and bundle equivalence.

We can summarize the relationship between linear bundles and affine bundles as follows.

- A linear vector bundle is an affine vector bundle endowed with a zero section.
- A linear principal bundle is an affine principal bundle endowed with a $R^n$-valued tensorial 1-form that projects to a zero section of an associated affine vector bundle.
- The tangent bundle (linear frame bundle) of a manifold $\Xi$ is a linear vector bundle (linear principal bundle) with
    - a solder form that identifies tangents to $\Xi$ and elements of the linear vector bundle, and
    - bundle transition maps defined by the coordinate transition maps of $\Xi$ and the solder form.

The last condition holds if and only if we can choose bundle a coordinate system so that the solder form is always represented by the Kronecker identity transformation. (Trautman 1973b).

If we start with an affine principal bundle that is not the affine frame bundle of $\Xi$, the local field theory over simply-connected regions is unaffected. However global invariants of the bundle, such as its characteristic classes, may enter into the theory.

## Contents